\documentclass[]{raa}                  
\usepackage{graphicx,times}             
\input{epsf.sty}                        
\input{psfig.sty}                       

\begin{document}

   \title{Heating Mechanisms and Radio Response from the Solar Chromosphere to Corona}

   \volnopage{Vol.25 (2025) No.0, 000--000}      
   \setcounter{page}{1}          

   \author{Baolin Tan \inst{1, 2} \and Jing Huang \inst{1, 2} \and Yin Zhang \inst{1}}
   \institute{State Key Laboratory of Solar Activity and Space Weather, National Astronomical Observatories, Chinese Academy of Sciences, Beijing 100012, China, {\it bltan@nao.cas.cn}\\
   \and School of Astronomy and Space Science, University of Chinese Academy of Sciences, Beijing 100049, China\\ }

   \date{Received~~2019 month day; accepted~~2019~~month day}

\abstract{Heating mechanism in the solar atmosphere (from chromosphere to corona) is one of the top-challenges in modern astronomy. The classic mechanisms can be divided into two categories: wave heating (W) and magnetic reconnection heating (X). Both of them still face some problems currently difficult to overcome. Recently, we proposed a new mechanism, called magnetic-gradient pumping heating (MGP, or P) which seems to overcome those difficulties, but still lacks sufficient observational evidence. Which one really explained the physics of hot corona exactly? How can observations be used to identify and verify the heating mechanism? Since different heating mechanism will generate non-thermal particles from different accelerations and experience different propagations, they will have different response on the broadband spectral radio observations. Among them, the non-thermal electrons from W mechanisms are closely related to shock-wave acceleration, and their radio response should be group of spike bursts with random distribution of drifting rates; the non-thermal electrons from X mechanisms are accelerated by reconnecting electric field with bidirectional flow, and their radio response should be type III pairs or spike pairs; P mechanism will produce energetic particle upflows, and their radio response should be unidirectional fiber bursts with moderate negative drifting rates. Therefore, the heating mechanism can be identified and verified from the the broadband dynamic spectral radio observations. Additionally, using high-resolution radioheliographs and spectral-imaging observations, the heating mechanisms in different regions can be identified and verified separately, thereby demonstrating the physical essence of hot corona.
 }

   \authorrunning{Tan, Huang, and Zhang}            
   \titlerunning{Coronal Heating and Radio Response}  

   \maketitle
\section{Introduction: Coronal Heating -- A Century long Scientific Challenge}
\label{sect:intro}

As is well known, all the energy released from the solar surface to the outside ultimately comes from the hydrogen nuclear fusions at the core of the Sun, where temperatures exceeds $1.5\times10^{7}$ K. After leaving that region, the temperature gradually decreases, down to 5700 K near the surface of the photosphere, and the photosphere also becomes weakly ionized plasma. But, more than 80 years ago, people were surprised to find that the outer atmosphere of the Sun (from chromosphere to corona) was much hotter than the inner photosphere, with the temperature increasing to beyond 10$^{6}$ K in the corona. If we do not doubt the correctness of the second law of thermodynamics, then there must be some nonthermal heating processes in the outer atmosphere of the Sun to maintain such high temperatures. What mechanism produces such heating processes? This is the mystery of coronal heating. Why do we have to study coronal heating? (1) The rapidly-warming transition region (TR) and the very hot corona with strong ionized, magnetic freezing plasmas is the source region of various solar activities through magnetic reconnections, naturally, the heating processes will dominate the generation of solar activities (including flares, coronal mass ejections, jets, etc.). (2) The hot TR and very hot corona are also the source of solar wind, the heating processes directly dominate the initiation, formation, and evolution of the solar wind. Therefore, elucidating the physical essence of solar atmospheric heating is an important prerequisite for us to understand the origins of solar activities, solar winds, and the occurrence of disastrous space weather events. (3) The coronal heating involves the generation, transportation, and dissipation of energetic particles in magnetized plasma, and this is a fundamental principle in plasma physics, which will provide a very important inspiration for studying the various instabilities in magnetic confinement nuclear fusion plasmas.

The heating energy requirement should be $\geq$300 $W m^{-2}$ above the solar quiet regions (QRs), $\geq$800 $W m^{-2}$ above the coronal holes (CHs), and $\geq 10^{4}$ $W m^{-2}$ above the active regions (ARs). For the chromosphere, due to its higher radiation loss, the heating energy requirement should be much higher, $\geq$4000 $W m^{-2}$ above the QRs and CHs, and $\geq 2\times10^{4}$ $W m^{-2}$ above ARs (Withbroe \& Noyes 1977). During the past more than 80 years, various heating mechanisms have been proposed to explain the formation of hot corona. The classic mechanisms can be simply classified into 2 kinds: wave heating (W mechanisms) and magnetic reconnection heating (X mechanisms) (Narain \& Ulmschneider 1996, Walsh \& Ireland 2003, and Klimchuk 2006, 2015, Leonardo 2023, etc.). However, each of them has some difficulties that are currently difficult to solve. In recent years, the heating effect of spicules and small-scale jets has also received increasing attentions (De Pontieu et al. 2004, 2011, Ji et al. 2012, Tian et al. 2013, Samanta et al. 2019, Chen et al. 2022, Chitta et al. 2023). Tan (2014) proposed a new mechanism, which is called magnetic-gradient pumping heating (MGP, or P mechanism in short). Although so much of heating mechanisms have been proposed, so far, it has not been fully solved, making it be a century long-lasting scientific problem (Kerr 2012). What is particularly confusing is that so far there is still no practical method to identify and verify different heating mechanisms through observations. These facts seriously constrain all our efforts to address this scientific challenge (Leonardo \& Fidel 2023).

Based on the differences in the generation and propagations of non-thermal energetic particles associated to different heating mechanisms, here, we discuss the possible physical image of plasma heating from the chromosphere to the corona, and the possibility of using radio broadband dynamic spectrum observations to identify and verify them. Section 2 discusses the main characteristics, generation and propagations of non-thermal energetic particles, and the main limitation in various heating mechanisms. Section 3 presents the possible patterns of radio responses associated with different mechanisms which may help us to identify and verify them. The conclusions are summarized in Section 4.

\section{Physical framework of heating mechanisms}

At first, we discuss the main characteristics, advantages, and main challenges faced by different heating mechanisms. Especially, we will focus on the generation and propagations of energetic particles in different heating processes, so that we may find out observable evidence to identify and verify the different heating mechanisms.

\subsection{The classic heating mechanisms}

Currently, the classical coronal heating mechanisms can be roughly divided into two categories: wave heating and magnetic reconnection heating. Here, we discuss their main physical processes and the basic characteristics of energy release separately.

\subsubsection{Wave heating (W mechanism)}

When the problem of coronal heating was first raised, the first mechanism that people thought of was wave heating. Due to the strong convection and turbulence in the photosphere, they can trigger the generation of various waves in the solar atmosphere, such as sound waves (Schwarzschild 1948), Alfven waves (Alfven 1947, Jess et al. 2009), and magnetoacoustic waves (Choudhuri et al. 1993), etc. If these waves can propagate upwards to the corona and dissipate there, there will be enough energy to heat the corona (Heyvaerts \& Priest 1983, Davila 1987, Lee \& Wu 2000, Cranmer et al. 2007, DePontieu et al. 2007, Ji et al. 2021, Yuan et al. 2023). Cranmer et al. (2007) developed a series of self-consistent models for the plasmas along open magnetic flux tubes rooted in CHs, streamers, and ARs to demonstrate the chromospheric heating driven by an empirically guided acoustic wave and the coronal heating from Alfven waves. Rappazzo et al. (2007) identified MHD anisotropic turbulence should be the physical mechanism responsible for the transport of energy from the large scales, where energy is injected by photospheric motions, to the small scales, where it is dissipated. Jess et al. (2009) reported the detection of Alfven waves in the low solar atmosphere with energy flux sufficient to heat the corona. Asgari-Targhi et al. (2013) and van Ballegooijen et al. (2017) constructed three-dimensional MHD models to simulate the spatial and temporal dependence of coronal loop heating by Alfven waves. However, it is still a serious question how does the wave energy dissipate into thermal energy in the surrounding plasmas?

One of the important wave dissipation is that in certain regions of the solar atmosphere, as the temperature and density of the plasma rapidly change, linear waves may transform into shock waves, and the shock wave may accelerate and generate non-thermal energetic particles, and therefore heat the plasma. This is the so-called linear wave - shock wave mechanism. The energy flux carried by a linear wave can be expressed as: $F=\frac{1}{2}\rho uV^{2}$. Here, $\rho$ is the plasma density, $V$ is the amplitude of disturbance velocity of the linear wave, and $u$ is the propagation speed, which is expressed as follows for sound waves ($v_{s}$), Alfven waves ($v_{A}$), and magnetoacoustic waves ($v_{m}$), respectively:

\begin{equation}
v_{s}=\sqrt{\frac{3k_{B}T}{m_{i}}},  v_{A}=\frac{B}{\sqrt{\mu_{0}\rho}},  v_{m}=[(v_{s}^{2}+v_{A}^{2})\frac{1\pm\alpha}{2}]^{1/2}
\end{equation}

Here, $k_{B}$ is the Boltzman constant, $T$ is the temperature, $m_{i}$ is the mass of ion, $B$ is the magnetic field strength, $\alpha=\sqrt{1-\frac{4v_{s}^{2}v_{A}^{2}\cos^{2}\theta}{(v_{s}^{2}+v_{A}^{2})^{2}}}$. $\theta$ is the angle between the propagation direction and magnetic field. In the expression of the magnetoacoustic waves ($v_{m}$), when + sign is taken, it is a fast magnetoacoustic wave, and when - sign is taken, it is a slow magnetoacoustic wave.

When a linear wave generating from the photospheric turbulence propagates upwards, its energy flux $F$ can be regarded as approximately conserved. Then, the amplitude of disturbance velocity can be expressed:

\begin{equation}
V\approx(\frac{2F}{\rho u})^{1/2}.
\end{equation}

From the photosphere to the corona, both the density and magnetic field strength decrease rapidly, while the temperature continues to increase. Overall, the value of $\rho u$ decreases, leading to an increase in the $V$ value. When $V>u$, the linear wave will transform into nonlinear sawtooth-like shock wave. We know that shock waves are nonlinear dissipative structures that can not only heat plasma, but also accelerate charged particles to form non-thermal energetic particle beams. Figure 1 presents the variations of propagating speed and disturbance velocity amplitude (purple dashed curves) with height above the solar photospheric surface for sound wave, Alfven wave, fast wave and slow wave. Here, the values of density and temperature come from the well-known literature (Vernazza et al. 1981), and the magnetic field strength ($B$) is obtained from potential field extrapolation with a simple assumption of 100 G at the photospheric footpoint ($B_{0}$), which is typical around the networks in QRs and CHs. For each type of wave, we provide results for three scenarios of energy flux: $F=$ 5, 50, and 500 $W m^{-2}$, respectively. Figure 1 may tell us the following facts:

(1) All linear waves originating from the photospheric convection are almost impossible to reach the corona, most of them transform into nonlinear shock wave in the chromosphere and the lower TR. Just because of this, non-thermal energetic particles associated with wave heating mechanisms should be mainly accelerated in the chromosphere and the lower TR.

(2) The larger the wave energy flux, the lower the height of upward propagation. Only when the energy flux is very low, can a small amount of waves penetrate through TR and reach to the corona. Obviously, this amount of wave energy is not enough to heat the corona.

\begin{figure}[ht]
\begin{center}
   \includegraphics[width=13 cm]{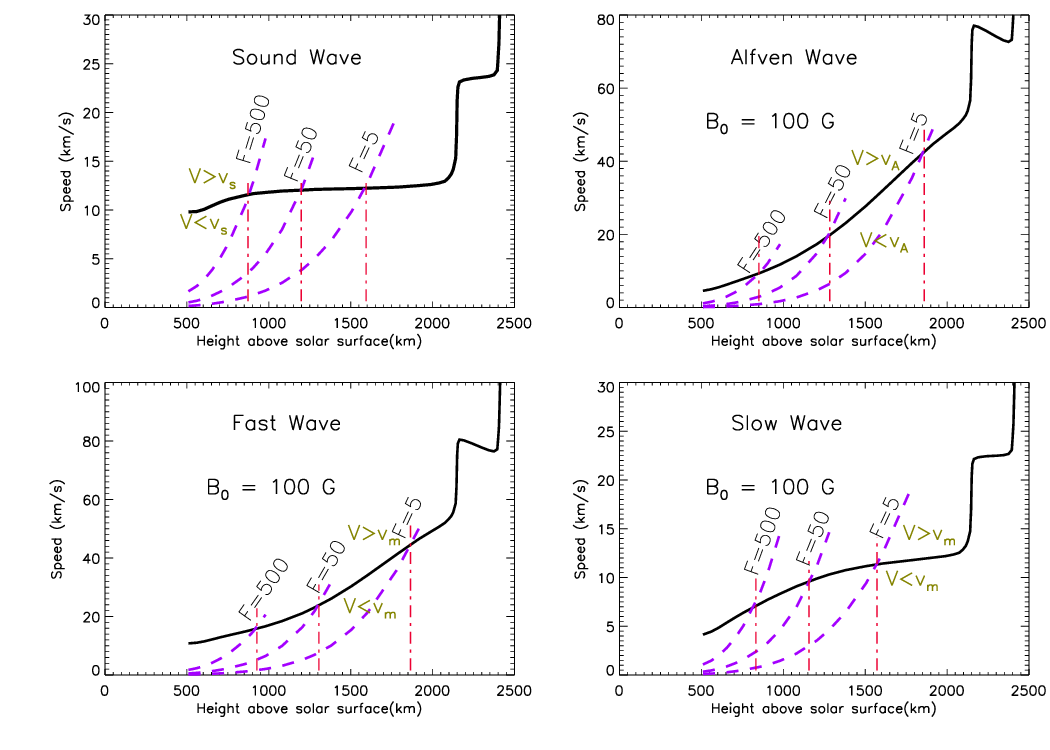}
\caption{The variations of propagating speed (black solid curves) and disturbance velocity amplitude (purple dashed curves) with height above the solar photospheric surface for sound wave, Alfven wave, fast wave and slow wave.}
\end{center}
\end{figure}

(3) We can also change the magnetic field $B_{0}$ in Figure 1, and it was found when $B_{0}$ increases, the height of the propagation of Alfven waves and fast magnetoacoustic waves also gradually increases, and the energy flux reaching the corona gradually increases. For example, wehn $B_{0}=1000$ G, $F=1$ $W m^{-2}$, the fast wave can propagate upward and reach to a height of 5000 km, closed to the bottom of the corona. As we know, the heating energy requirement should be $\geq 10^{4}$ $W m^{-2}$ in the corona above ARs. This indicates that in AR, only a very small part of linear waves originating from the photosphere can propagate through into the corona before being converted into shock waves and dissipated.

However, there are also many observations indicating the existence of various waves in the corona (Van Doorsselaere et al. 2020, Hashim et al. 2021). There are two possibilities for this, one is that these waves come from leakage in a few very strong magnetic field regions which should be only a tiny fraction of contribution to heat the corona. The other is that these waves do not originate from photospheric convection, but are triggered by various eruptions in the corona, such as flares, microflares, or even nanoflares. Actually, the heating of the corona by such waves is part of the magnetic reconnection heating, which will be discussed in the following section.

The above facts indicate that whether it is sound waves, magnetoacoustic waves, or Alfven waves, their heating effect is limited to the solar chromosphere and low TR, and basically ineffective to the higher solar atmosphere.

\subsubsection{Magnetic reconnection heating (X mechanisms)}

As we often see, solar flares can release a large amount of energy into the surrounding plasma through magnetic reconnection. For example, Lu et al. (2024) found that the continuous emergence of magnetic flux in active regions might drive magnetic reconnections that release energy impulsively but persistent over time on average. As a result, numerous substructures are heated to 10$^{7}$ K. Besides the powerful flares, Jin et al. (2021) reported a micro-flare event on the quiet Sun, which release 1.3$\times10^{27}$ erg of thermal energy and heated the corona up to 5.8$\times10^{6}$ K. Perhaps we can assume that there are a series of flares of various scales in the solar atmosphere, including the powerful X-class flares, M-class flares, C-class flares, and the small B-class flares, A-class flares, and even microflares, nanoflares, and picoflares. The energy released by them can heat the solar chromosphere, TR, and corona. Although the scales of these flares vary greatly, their released energy is essentially magnetic reconnection around the current sheet in the source region, which accelerates the charged particles and heats up the surrounding plasma. This heating mechanism is also known as magnetic reconnection heating mechanism, or DC mechanisms, or nanoflare models (Parker 1988, Hudson 1991, Sturrock 1999), here, we marked it as X mechanism. Many people have studied the possibility of this mechanism heating the corona through observations and numerical simulations (Testa et al. 2014, Tian et al. 2014, Bradshaw \& Klimchuk 2015, Berghmans et al. 2021, Chen et al. 2021).

If the energy released by each micro-flare is approximately on the order of $10^{19}$ J, the occurrence rate of micro-flares on the whole solar surface for heating the corona must be $>1000 s^{-1}$ (Moore et al. 1991). However, so far, all statistical observations indicate that the actual occurrence rate of micro-flares is less than 10 $s^{-1}$ (Hudson 1991, Jiang et al. 2015). This fact indicates that X mechanisms likely provide only a small part of contribution to coronal heating. Additionally, the relation between the frequency of the events $f(E)$ and their energy $E$ follows a power law,

\begin{equation}
f(E)=f_{0}E^{-\alpha}.
\end{equation}

It has long been established that the power law index $\alpha$ is a strong indicator of whether X mechanism is an important coronal heating mechanism. If X mechanism is enough to provide a sufficient contribution to heat the corona, it requires $\alpha>2$ (Hudson 1991, Aschwanden \& Freeland 2012). However, Crosby et al. (1993) obtained $\alpha=1.53\pm0.02$ for events with energies in the range $10^{21}$ - $10^{24}$ J. Shimizu (1995) estimated $\alpha$ to be between 1.5 to 1.6 for the active-region small-scale events with energies in the range of $10^{20}$ - $10^{22}$ J. These results seem to indicate that X mechanism cannot be the dominant heating mechanism in the solar corona. However, on the contrary, optimistic results have also been reported. For example, Parnell \& Jupp (2000) obtained a result of $\alpha>2.0$ events with energies in the range of $10^{16}$ - $10^{19}$ J, and implied that X mechanism may dominate the heating of the solar quiet corona. These facts have once again puzzled us, how much does the X mechanism contribute to coronal heating?

With more and more high-resolution imaging observation equipment put into operation, a huge amount of observation data has been accumulated, and people are constantly conducting more extensive and in-depth statistical research, hoping to obtain more accurate conclusions. A very recent statistical study is showed $\alpha=1.63\pm0.03$ (Mason et al. 2023), which indicated that the contribution of X mechanism to coronal heating is almost less than 20\%. More than 80\% of the required energy for coronal heating should still be provided by W or other heating mechanisms. We believe that this latest statistical result may be closer to the truth about the Sun, and the X mechanism may indeed only play a minor role in the heating process of the entire solar atmosphere.

In addition, since neutral particles have no response to magnetic fields, magnetic reconnection is not easy to occur in the low chromosphere with weak ionization at low temperatures. Correspondingly, X heating is difficult to occur in the low chromosphere. Only in strong or fully ionized plasma of upper chromosphere, TR, and corona can X heating occur.

\subsection{New heating mechanism}

Is there other heating mechanism besides W and X mechanisms? In recent years, new mechanisms or ideas have been constantly proposed, such as spicules, small-scale jets, EUV cyclones, and magnetic-gradient pumping mechanism, etc.

\subsubsection{Spicules, Small-scale jets, and EUV cyclones}

Recently, with a series of high-resolution solar telescopes put into operation for observation, more and more characteristics related to spicules and small-scale jets have been discovered. Their heating effect on the chromosphere and corona has also received increasing attentions, and people even suspect that this may be a new kind of heating processes (De Pontieu et al. 2004, 2011, Tian et al. 2013, Samanta et al. 2019, Chitta et al. 2023). De Pontieu et al. (2004) presented that the p-modes might leak suf?cient energy from the global resonant cavity into the chromosphere to power shocks and drive upward ?ows to form spicules. Furthermore, De Pontieu et al. (2011) revealed a ubiquitous coronal mass supply from the chromospheric plasma in fountain-like jets or spicules to accelerate upward and may heat the corona. Zhang \& Liu (2011) demonstrated that the ubiquitous EUV cyclones rooted in the rotating network magnetic fields in the quiet Sun might provide an effective way to heat the corona. Ji et al. (2012) revealed the unexpected complexes of ultra-fine, hot magnetic channels linking from the photosphere to the base of the corona with upward hot plasma jets.

Physically, when we carefully analyze the above observational features, we find that most of the spicules and small-scale jets are actually still closely related to certain magnetic reconnection or cancellation, and they are likely a new manifestation of X heating (Tian et al. 2013, Samanta et al. 2019, Chitta et al. 2023). There are also some events of spicules and small-scale jets are related to certain wave processes originating from photospheres (De Pontieu et al. 2004, Cranmer \& Woolsey 2015). Moreover, utilizing the MGP mechanism that we will discuss in the next section can provide a reasonable explanation for some spicules, jets, and ultrafine hot channels. Therefore, although these small-scale activity events do contribute to heat the solar atmosphere, they do not physically reflect an independent heating mechanism.

\subsubsection{Magnetic-gradient pumping mechanism (P mechanism)}

In fact, magnetic gradients are commonly present in the solar atmosphere above ARs, magnetic networks in QRs, and even CHs, and in this case, the magnetic-gradient force will also drive the aggregation of energetic particles in the weak field region (Tan 2014, Tan et al. 2020).

\begin{equation}
F_{t}=mg(h)-G_{B}\epsilon_{t} ,
\end{equation}

Here, $G_{B}=\frac{\bigtriangledown B}{B}$ is the relative magnetic gradient, $\epsilon_{t}$ is the transverse kinetic energy of the charged particles. $mg(h)$ is the gravitational force at height of $h$ above solar surface, while $G_{B}\epsilon_{t}$ is the magnetic-gradient force. Due to the fact that the magnetic field strength in the solar atmosphere always decreases with height $h$, the direction of the magnetic gradient ($G_{B}$) is always downward, while the direction of the magnetic-gradient force is always upward. This feature is independent of the direction of magnetic fields themselves.

When $F_{t}=0$, $\epsilon_{t}=\frac{mg(h)}{G_{B}}=\epsilon_{0}(h)$, the particle will reside around the height of $h$, called resident particles. $\epsilon_{0}(h)$ is a critical kinetic energy, which is a function of height ($h$). At each height $h$, the temperature is dominated by $\epsilon_{0}(h)$ which is determined by the local gravitational force $mg(h)$ and the relative magnetic gradient $G_{B}$. Generally, the relative magnetic gradient $G_{B}$ is always slowly decreasing in the chromosphere, rapidly decreasing in TR, and then becoming slowly decreasing in the corona, and just because of this variation, the temperature become slowly increasing in the chromosphere, rapidly increasing in TR, and slowly increasing in the corona. In addition, as $G_{B}$ decreases with increasing height and decreases faster than the gravity $mg(h)$, temperature $T$ increases with increasing height. The above variation just coincides perfectly with the actual situation in the solar atmosphere. According to this principle, it can precisely explain the formation of type II spicules (De Pontieu et al. 2011) and the ultra?ne hot channels (Ji et al. 2012).

When $F_{t}<0$, $\epsilon_{t}<\epsilon_{0}(h)$, the particle will sink down to below the height of $h$, called confined particles. This indicates that lower-energy particles will always stop and remain in the lower atmosphere.

When $F_{t}>0$, $\epsilon_{t}>\epsilon_{0}(h)$, the particle will move upward to escape from the height of $h$, called escaping particles. This indicates that higher-energy particles will be pumped upwards to form energetic particles upflows and carry energy to the higher atmosphere.

Essentially, the MGP heating mechanism belongs to a sorting process without particle acceleration and therefore without magnetic energy release. The magnetic gradient separates energetic particles from the lower atmospheric plasma with thermal equilibrium and drives them to move upwards. The upward escaping energetic particles accumulate in the upper atmosphere and thereby increasing the temperature of the upper atmosphere. This process is somewhat similar to mineral processing: a small amount of rich ore is extracted from a large amount of poor ore, and the total ore quantity is conserved. The primary estimation indicates that the energy carried by the upwards escaping energetic particles is sufficient to heat and maintain the hot TR and the hotter corona (Tan 2014). By utilizing the MGP mechanism, perhaps we may provide a reasonable explanation for the formation of the ubiquitous type II spicules (De Pontieu et al. 2011).

In addition, for fully ionized magnetized plasmas, the MGP mechanism can play a very natural role. However, for partially ionized or even weakly ionized plasmas, such as chromospheric plasmas, the collision process can easily cancel out the MGP process due to the presence of a large number of neutral particles. Therefore, the optimal regions for the MGP mechanism to work are the highly ionized upper chromosphere, transition region, and corona.

\subsection{Overall physical image of heating from the chromosphere to corona}

The above three kinds of heating mechanisms all depend on the magnetic field from the chromosphere to the corona, but their dependence on the magnetic field is different, and the corresponding heating process is also completely different. For W heating, the magnetic field provides a waveguide, but for wave dissipation, the role of the magnetic field varies in different dissipation mechanisms. For X heating, the changes of magnetic field topology is a fundamental process for magnetic energy release. In P mechanism, the magnetic field provides a pumping channel, and the whole heating process does not change the strength and configuration of the magnetic field, nor does it release magnetic energy. Instead, the magnetic-gradient force redistributes the thermal particles in the lower atmosphere in space according to their energy, and high-energy particles converge towards the region with weak magnetic field in the corona. The above differences also indicate that different heating mechanisms can be applied to different physical environments. In fact, the differences in physical conditions from the chromosphere, transition region to the corona of the sun are very significant. As shown in Figure 2, the chromosphere has a slowly rising temperature, and the gas is weakly ionized, with ionization degrees ranging from a negligible 0.01\%  to about 10\%. TR has a rapidly increasing temperature, and the gas is basically strongly ionized. The corona's temperature becomes rising slowly again, and the gas becomes a very thin fully ionized plasma. These facts imply that the coronal heating problem is actually composed of three closely related and interdependent sub-problems, namely chromospheric heating, TR heating, and coronal heating.

\begin{figure}[ht]   
\begin{center}
   \includegraphics[width=12 cm]{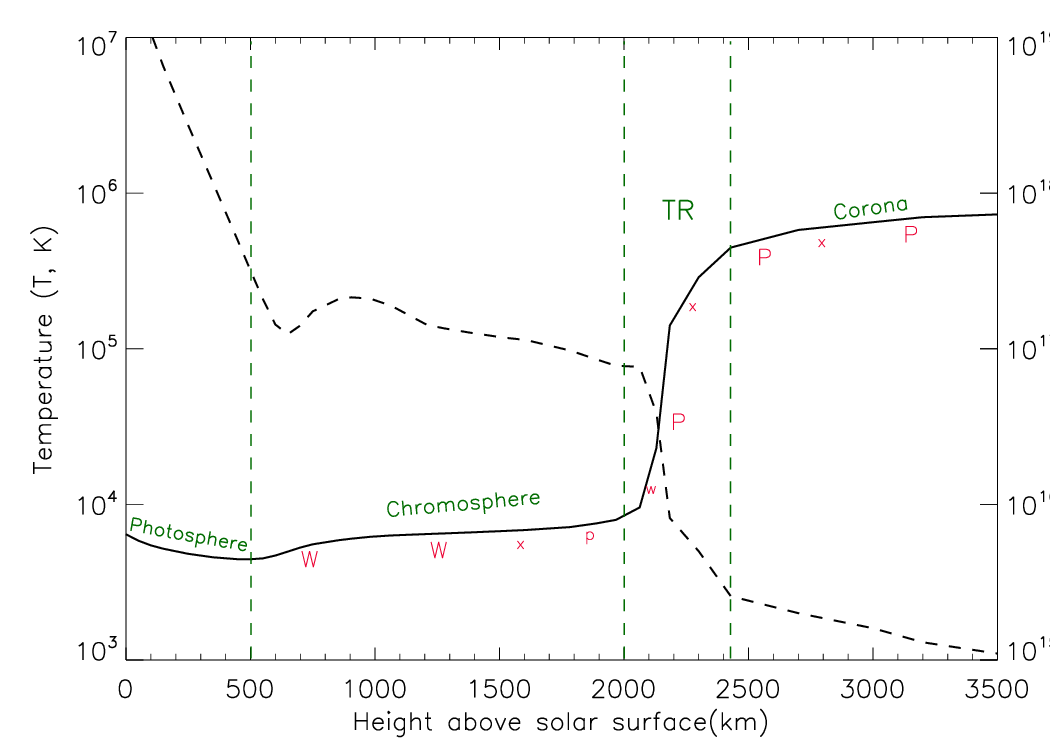}
\caption{The variations of temperature (solid) and electron density (dashed) of the photosphere, chromosphere, TR, and corona above the solar surface as a function of height. The letters marked below the temperature curve indicate the possible heating mechanisms in different regions, with uppercase letters indicating the dominant heating mechanism and lowercase letters indicating the partially contributing heating mechanism. W represents wave heating, x represents reconnection heating, and P represents MGP heating.}
\end{center}
\end{figure}

Based on the above discussion, we attempt to establish the following physical image of solar atmospheric heating:

(1) In chromosphere, the photospheric convection can excite various waves, including sound waves, magnetoacoustic waves, and Alfven waves, and transport energy upwards. As we discussed in section 2.1.1, these waves are difficult to penetrate through TR and reach to the corona. In addition, the weak partial ionization of the dense low chromosphere make it difficult for the X and P heating processes to work. Therefore, we can preliminarily infer that various waves (marked as $W$ in Figure 2) originating from the photospheric convection may be dominantly heat the chromosphere, result in the slowly increasing of temperature. It is likely that in the upper chromosphere, with the temperature increases to beyond $10^{4}$ K and the ionization degree becomes high enough and makes X heating and P heating to provide a certain contribution (marked as $x$ and $p$ in Figure 2).

(2) In TR, due to the strong ionization and strong magnetic fields, P heating should be dominant (marked as $P$ in Figure 2) and W heating can provide a minor contribution in the lower TR. Various scales of X heating (small flares, microflares, nanoflares, etc.) also provide some contributions here.

(3) In the corona, the gas becomes fully ionization, and magnetic field becomes relatively weak, the dominant heating mechanism should be P heating, and X heating (flares, small flares, microflares, etc.) provides a minor contribution, and W heating can be neglected.

In Figure 2, we have marked the possible heating mechanisms in different regions below the temperature curve with letters, where uppercase letters ($W$, $P$) indicate the dominant mechanism and lowercase letters ($w$, $x$, $p$) indicate the secondary heating mechanism. Here, $W$ and $w$ marked the wave heating mechanism, $P$ and $p$ marked the MGP heating mechanism, and $x$ marked magnetic reconnection heating mechanism.

\section{Radio emission responses in different heating mechanism}

In the above section, we have shown that there are various heating mechanisms (W, X, and P) and different combinations in different regions, then, how do we distinguish and identify them from observations? As we know that any heating mechanism should include two aspects: thermal (temperature increasing) and nonthermal effects (energetic particles generating and propagating). In different heating mechanisms, the observed characteristics of the thermal process are almost similar to each other (such as images at infrared, optical, UV and EUV wavelengths), making it difficult to distinguish from observations. However, non-thermal processes are significantly different. Due to such significant differences in the generation and propagation of energetic charged particle flows from different heating mechanisms, and inevitably producing vastly different emission signals in radio observations, making it possible to distinguish different heating mechanisms accordingly. Radio emission has a clear response to almost all processes in astro-plasmas, including thermal phenomena, non-thermal phenomena, and the variations of magnetic field. Especially, radio emission is very sensitive to the non-thermal energetic electrons. Different particle accelerations and propagations have different responses in radio broadband dynamic spectral observations (Dulk 1985, Bastian et al. 1998, Gary 2023). As we know that solar radio emissions cover a wide frequency range over up to 7 orders of magnitude from submillimeter waves ($>1$ THz) to kilometer waves ($<300$ kHz), and the corresponding source region cover the solar photosphere, chromosphere, TR, corona, and even the vast interplanetary space, and the emission mechanisms include the bremsstrahlung and cyclotron emission (CE) of the thermal or low-energy nonthermal electrons, the gyrosynchrotron emission (GE) and coherent plasma emission (PE) of the nonthermal electrons. As discussed in Section 2, the generation of energetic particles varies in different coronal heating mechanisms, which may exhibit different spectral patterns in radio observations, including their duration, bandwidth, lifetime of single burst, frequency drifting rate and its distribution of positive and negative signs, etc.

\subsection{Radio response of W mechanisms}

When W mechanisms play main role in coronal heating, the linear MHD waves (sound waves, Alfven waves, fast and slow magnetoacoustic waves) have to be transformed into shock waves in order to dissipate energy and heat the plasma. It is just because of the appearance of shock waves that charged particles have the opportunity to be accelerated, producing non-thermal energetic particles. The shock wave acceleration of charged particles often has a very complex randomness, which results in randomness in the propagation direction of the generated non-thermal particles after leaving the accelerating site, and the corresponding radio emission should be a large group of spike bursts with short lifetime ($<$ 1 ms), narrow frequency bandwidth (around 1\% of the central frequency), and random distribution of positive and negative signs of fast frequency drift rate, similar to the spike groups associated with flare terminal shock waves, or in short (Chen et al. 2015). The left panel of Figure 3 shows an example of randomly distributing spike group (RDSG), which took place in the decay phase of a powerful X-class flare on 2005 Jan 20.

\begin{figure}[ht]   
\begin{center}
   \includegraphics[width=15 cm]{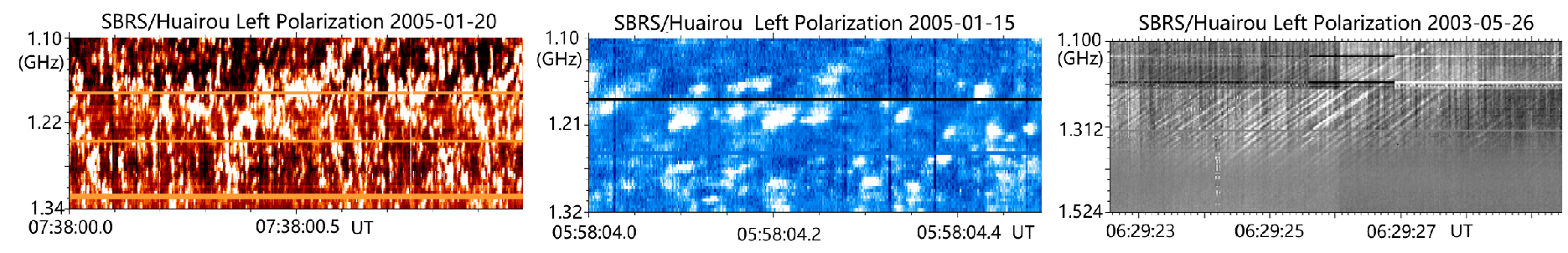}
\caption{Possible radio response signals of the non-thermal processes with different heating mechanisms. The left shows a randomly distributed spike group that may be related to W heating, the middle shows spike pairs that may be related to X heating, and the right shows a fiber burst group that may be related to the P mechanism.}
\end{center}
\end{figure}

As we discussed in Section 2.1, the various waves originating from photosphere convection are difficult to penetrate through TR, and most of them transform into shock waves in the chromosphere and lower TR. Therefore, the non-thermal electrons associated with W heating generate and propagate mainly in the chromosphere and lower TR. They can generate radio emission in two ways: GE and PE (Dulk 1985). In the chromosphere, the electron density ($n_{e}$) is between $10^{17}$ $m^{-3}$ to $4\times10^{17}$ $m^{-3}$ (Vernazza et al. 1981, Figure 2), and the magnetic field strength can be assumed to be ranged from 100 G to 1000 G. The nonthermal electrons can be regarded as mildly relativistic with kinetic energy of 10 - 300 keV. Then, the associated PE (include fundamental and second harmonic) should be in the frequency range of 2.8 - 11.5 GHz, and GE with harmonic number of 10 - 100 will be in the frequency range of 2.8 - 280 GHz. In the transition region, the electron density ($n_{e}$) is between $2\times10^{15}$ $m^{-3}$ to $10^{17}$ $m^{-3}$ (Figure 2), the magnetic field strength will be slightly weaker than in the chromosphere and can be assumed that the typical values is about 50 - 500 G. With these parameters, the associated PE will be in the frequency range of 400 MHz - 5.7 GHz, and GE should be in the frequency range of 1.4 - 140 GHz. In fact, in the chromosphere and TR, there are not many non-thermal electrons with energies exceeding 300 keV, and the corresponding emission is also weak. The harmonic number of GE is difficult to exceed 70 - 80, and the corresponding upper limit of the emission frequency is always lower than 100 GHz. Due to the lower electron density and weaker magnetic field in the corona compared to TR, if we can observe RDSG below the above frequency range during the quiet Sun, it indicates that W heating also exists in the corona.

\subsection{Radio response of X mechanisms}

When X mechanism plays main role for heating the corona, then there will inevitably be a large number of small-scale current sheets in TR and corona. Here, the energetic non-thermal particles are accelerated through the reconnecting electric field, and form many bidirectional non-thermal particle flows (Yu et al. 2020), and the corresponding radio emission will have a bidirectional fast drifting spectral structure appearing in burst pairs with opposite frequency drifting rates, such as type III pairs (Aschwanden \& Benz 1997, Tan et al. 2016), or spike pairs (Tan 2013). In each burst pair, the separation frequency between the positive and negative drifting branches indicates the position of the corresponding site of magnetic reconnection and particle acceleration. The middle panel of Figure 3 shows an example of spike pairs, which took place around an M8.6 flare on 2005 Jan 15.

In the process of magnetic reconnection, a part of non-potential magnetic energy is directly converted into thermal energy and heat the surrounding plasma, while the other part of non-potential magnetic energy accelerates charged particles to generate non-thermal energetic particles by the reconnecting electric fields around the current sheets, and the energy of the non-thermal electrons ($E_{acc}$, unit of keV) can be estimated from the following expression (Singh 2015, Tan et al. 2024),

\begin{equation}
E_{acc}\approx1.23\times10^{13}\frac{B^{2}}{n_{e}} ,
\end{equation}

Here, $B$ is the magnetic field strength (unit of G) near the reconnecting site and $n_{e}$ is the electron density ($m^{-3}$). In the chromosphere, typically $n_{e}=10^{17}$ $m^{-3}$, $B=1000$ G, then $E_{acc}\approx 120$ keV. In TR, $n_{e}=10^{16}$ $m^{-3}$, $B=500$ G, then $E_{acc}\approx 300$ keV. And in the corona, $n_{e}=10^{15}$ $m^{-3}$, $B=100$ G, then $E_{acc}\approx 120$ keV. In brief, the energy of the non-thermal electrons accelerated by the reconnecting electric field may beyond 100 keV and the electrons become mildly relativistic. These non-thermal electrons propagate along the direction perpendicular to the magnetic field, usually forming bidirectional outflow beams. The associated radio emission should be generated from GE with harmonic number of 10 - 100 and PE (fundamental and second harmonic). Similar to the discussion in Section 3.1, in the chromosphere, the related frequency of PE should be in the range of 2.8 - 11.4 GHz and GE should be in the range of 2.8 - 280 GHz. In TR, the corresponding frequency range will be 400 MHz - 5.7 GHz for PE and 1.4 - 140 GHz for GE (the actual upper limit of the emission frequency is difficult to exceed 100 GHz in the chromosphere and TR). As for lower corona, the typical the electron density can be assumed $10^{15}$ $m^{-3}$, and the typical magnetic field strength 10 - 100 G. Then the associated PE will be in the frequency range of 280 - 560 MHz, and the associated GE will be in the frequency of 280 MHz - 5.6 GHz.

As presented in Section 2.1, the various MHD waves generated from the photosphere are difficult to penetrate through TR and reach to the corona, so even if W heating occurs in the lower solar atmosphere, it is still difficult to occur in the corona above TR. But for X mechanism, the current sheets and the corresponding magnetic reconnections may appear in any regions with strong magnetic-freezing plasmas from the upper chromosphere, transition region to the corona. The corresponding radio burst pairs will occur in a wide frequency range from 280 MHz to 280 GHz, where the burst pairs in the mm-cm wavelength band are likely a signal of X heating in TR, while the burst pairs in the dm-m wavelength should be a signal of X heating in the corona.

\subsection{Radio response of P mechanisms}

When MGP mechanism (marked as P) operates in the solar atmosphere, then energetic particle upflows will be generated in the upper chromosphere, TR, and corona. The critical energy $\epsilon_{0}(h)$ determines the lower limit of kinetic energy and velocity of the energetic particle upflow at height $h$. By averaging all escaping particles, the number, average velocity and energy of the upflows can be obtained (Tan 2014). And the calculations show that the averaged energy of escaping particles ranges from about 10 eV to 100 eV in TR, and from 100 eV to several keV in the corona (Tan 2014). The corresponding velocity of the upflows will range from about 2000 -- 1.0$\times10^{4}$ km$\cdot s^{-1}$ in TR, and (1.0 -- 3.0)$\times10^{4}$ km$\cdot s^{-1}$ in the corona. This velocity is much higher than the local Alfven speed (about 100 -- 500 km$\cdot s^{-1}$ in TR, and 500 -- 2000 km$\cdot s^{-1}$ in the corona) and much smaller than that of the non-thermal particles accelerated by shock wave or magnetic reconnection (generally, the velocity of non-thermal particles is $>0.2$c, that is $>6.0\times10^{4}$ km$\cdot s^{-1}$, and the corresponding energy is $>10$ keV), but it is still significantly much higher than the local electron thermal velocity in TR (about 400 -- 2000 km$\cdot s^{-1}$ and in the corona (2000 -- 4000 km s$^{-1}$) and Alfven speed in TR (about 200 -- 600 km$\cdot s^{-1}$) and in the corona (600 -- 1500 km s$^{-1}$). These unidirectional energetic particle upflows are sufficient to excite Langmuir waves in plasma and generate PE in frequency range of 2.8 - 11.4 GHz in the chromosphere, 400 MHz - 5.7 GHz in TR, and 280 MHz - 2.8 GHz in the corona. The related radio emission will have intermediate negative frequency drifting rates which is much faster than type II radio bursts and much slower than type III radio bursts. They are somewhat similar to the radio fiber bursts (Wan et al. 2021). The right panel of Figure 3 presents a group of fiber bursts, which took place far from the peak phase of an M1.9 flare. Naturally, the energetic electron upflows triggered by MGP mechanism can also generate radio signals through CE under the action of a magnetic field, with corresponding harmonic numbers of 3 - 10 and the corresponding frequencies of 840 MHz - 28 GHz in the chromosphere, 420 MHz - 14 GHz in TR, and 100 MHz - 2.8 GHz in the corona.

As we know, the fast drifting radio bursts, such as type III radio bursts, spike bursts, are triggered by nonthermal high-energy electron beams with velocity of $>0.2$c propagating in plasmas (Reid \& Ratcliffe 2014, Tan et al. 2019), while the slow drifting radio bursts, such as type II radio bursts are triggered by coronal mass ejections (CMEs) or jets with velocity $<2000$ km$\cdot s^{-1}$ (Wager \& MacQueen 1983, Su et al. 2015, Hou et al. 2023). Then, what triggers the fiber bursts with intermediate frequency drifting rates? The formation mechanism of solar radio fiber bursts has not been well explained (Alissandrakis et al. 2019, Bouratzis et al. 2019). Based on the above discussions, we propose that the energetic particle upflows formed by MGP mechanism may precisely trigger the formation of fiber bursts. Perhaps this is a more natural and reasonable explanation for the cause of solar radio fiber bursts with intermediate frequency drifting rates.

\subsection{How to observe the radio response signals of solar atmospheric heating?}

From the above discussion, we may find that, due to the different generations and propagations of energetic particle flows in different heating mechanisms, the spectral characteristics of the excited radio emission vary greatly. These differences are summarized in Table 1. These radio emission occurs typically in a wide frequency range of 100 MHz - 100 GHz (2.8 - 280 GHz in the chromosphere, 400 MHz - 100 GHz in TR, and 100 MHz - 28 GHz in the corona). Here,  the frequencies corresponding to different emission mechanisms also vary slightly. That is to say, in order to verify the heating mechanism in different regions of the solar atmosphere, we need dynamic spectral observations in a wide frequency range of 100 MHz - 100 GHz, which covers meter wave, decimeter wave, centimeter wave, till to millimeter waves. If we can observe different regions (AR, QR, CH, or polar region) of the Sun through ultra-wide frequency band radio spectrometers and extract the spectral parameters (including the frequency bandwidth, lifetime, frequency drifting rate, polarization degree, repetition rate and group distribution) from them, it is possible to identify and verify their heating mechanisms.

\begin{table}
\begin{center}
 \caption[]{Summary of radio responses among different heating mechanisms in solar atmosphere.}
 \begin{tabular}{ccccccccccc}
  \hline\noalign{\smallskip}
                  & W mechanisms    & X mechanisms                   & P mechanism          \\\hline\noalign{\smallskip}
Magnetic field    & wave guide      & energy release                 & pumping channel         \\
Acceleration      & shock wave      & reconnecting electric field    & magnetic-gradient pumping \\
Propagation       & random flows    & bidirectional flows            & unidirectional upflows  \\
Particle energy   & 10-1000 keV     & 10-1000 keV                    & 20 eV - 2 keV         \\
Radio emission    & spike groups    & type III/spike pairs           & fiber groups            \\
Freq. drifting rate & fast          & fast                           & moderate                \\
Sign of drifting rate & random      & Positive/negative pairs        & negative                \\
Freq. range       & 280 MHz - 100 GHz  & 280 MHz - 100 GHz           & 100 MHz - 28 GHz        \\
\noalign{\smallskip}\hline
\end{tabular}
\end{center}
\end{table}

From the above discussions, we also found that the frequency of the solar atmospheric heating non-thermal radio emission is not only related to the atmospheric layers (the chromosphere, TR, or corona), but also to the emission mechanism, which may result in radio response signals from the different layers appearing in the same frequency range. How can we distinguish it?

We know that the frequency of the coherent PE is: $f = 9sn_{e}^{1/2}$. Here, $s=1$ is the fundamental plasma emission, $s=2$ is the second harmonic plasma emission. The corresponding frequency drifting rate is: $\frac{df}{dt}=\frac{9s}{2}n_{e}^{-1/2}\frac{dn_{e}}{dr}v_{e}$. That is to say,  $\frac{df}{dt}\propto\frac{dn_{e}}{dr}$. Here, $v_{e}$ is the velocity of the energetic electrons. From Figure 2, it can be seen that in both the chromosphere and corona, the electron number density decreases slowly, and correspondingly, its gradient $\frac{dn_{e}}{dr}$ is relatively small; while in TR, the gradient $\frac{dn_{e}}{dr}$ rapidly decreases, and its gradient is very large. This also indicates that the frequency drifting rate of non-thermal radio emission occurring in the chromosphere and corona is relatively small, while the frequency drifting rate in TR is very high. This can serve as an important criterion for distinguishing the non-thermal radio signals from the chromosphere, TR, and corona. As for the incoherent CE (low-energy energetic upflow electrons produced by P mechanism) and GE (mildly relativistic electrons produced by W and X mechanism), $f = sf_{ce}$, here, $f_{ce}$ is the electron gyro-frequency and $s$ is the harmonic number. The related frequency drifting rate is proportional to the magnetic gradient: $\frac{df}{dt}\propto\frac{dB}{dr}v_{e}$. Similar to the characteristics of density changes, the magnetic field in the chromosphere and corona gradually decreases with a small gradient, while in TR, it rapidly decreases with the highest gradient. This characteristic of change determines that the frequency drifting rate is relatively small in the chromosphere and corona, and highest in TR.

We know that radio emissions with frequencies below 15 GHz are completely transparent to the Earth's atmosphere and can be directly observed using ground-based solar radio telescopes. However, unlike radio telescopes specifically designed to observe solar eruptive activities (such as flares and CMEs), the observation targets here are mainly non-thermal emission signals during the quiet Sun, including spike groups, spike pairs or type-III pairs, fiber groups, etc. They require telescopes to have very high sensitivity, high time-resolution and frequency-resolutions, which are difficult for general radio telescopes to accurately detect above signals. Only large aperture telescopes or telescope arrays composed of multiple elements can have the ability to detect the above signals, such as MUSER (40/60 dishes, 400 MHz - 15.0 GHz, covers from the upper TR to corona, and is very suitable for detecting the radio bursts coming from ARs, Yan et al. 2021), FAST Core Array (0.35 - 10.0 GHz, covers from the upper TR to corona and with very high sensitivity and high spatial resolution, and is very suitable for detecting weak signals of the radio bursts coming from the quiet Sun and ARs. Jiang et al. 2024), and the future SKA, etc.

Due to the strong absorptions of Oxygen and water molecules in the Earth's atmosphere, so far, there is currently no existing telescope capable of obtaining broadband dynamic spectrum observations typically in frequency range of 15 - 100 GHz. At present, the broadband dynamic spectrum observation in this frequency range is still blank both domestically and internationally. Therefore, in order to identify and verify the different heating mechanism in the solar atmosphere, we expect to implement a radio exploration plan for the heating mechanism of the chromosphere and transition region through the following three steps. The first step is to select frequency bands with weaker absorption in the Earth's atmosphere and develop new broadband dynamic spectrometers in dry and high-altitude place, such as the Solar Ultra-Broadband Millimeter-wave Spectrometers (SUBMS) currently under development and construction at the Lenghu Station in Qinghai Province, with observation frequency range of 15 - 36 GHz (Tan et al. 2024). SUBMS may provide the information of the non-thermal heating processes in the upper transition region. After this step is successfully implemented, we will consider to implement the second step, that is to onload a radio dynamic spectrometers on a certain space platform to observe in the full frequency range (15 - 100 GHz, cover the chromosphere and transition region) from outside of the Earth's atmosphere. This step will obtain clean broadband dynamic spectrum with high temporal - spectral resolutions. When the second step is also successful, we will consider the third step, which is to build a millimeter wave telescope array on a platform such as the Chinese Space Station, and use the synthetic aperture principle to carry out broadband dynamic spectral imaging observations, detecting heating signals in the solar atmosphere above ARs, QRs, CHs, and polar region, respectively.

Through the above joint observations of ground-based and space telescopes in the frequency range from millimeter wave to meter wave, we may obtain complete information on the heating process from the solar chromosphere, TR to the corona, providing observational evidence for solving the mystery of coronal heating.

\section{Conclusion}

Coronal heating is a very difficult and complicated problem. Based on the above discussions, we obtained the following conclusions:

1. It is likely that multiple heating mechanisms (W, X, and P heating mechanism) working together in the solar atmosphere, and the contribution of different mechanism may be different from the chromosphere, TR, to corona. W mechanism may dominate the heating process in the chromosphere and low part of TR, while P mechanism may dominate the heating processes in TR and the corona, and X heating is likely to provide only a small contribution to heat the solar atmosphere. Spicules and small-scale jets which have been widely discussed in recent years are partly related to X heating, partly related to W heating, and partly formed from P processes.

2. Among these heating mechanisms, the generation and propagation characteristics of non-thermal or energetic electrons are different, they produce different spectral structural features on the radio broadband dynamic spectrums: W heating is mainly manifested as RDSG, X heating is manifested as type III burst pairs or spike pairs, and P mechanism is manifested as fiber burst groups. The possible related emission mechanism includes coherent PE, incoherent CE (low-energy energetic upflow electrons produced by P mechanism) and GE (mildly relativistic electrons produced by W or X mechanisms). For each kind of radio bursts, when it comes from the chromosphere, the frequency drifting rate generally should be relatively slow in the high frequency range (2.8 - 280 GHz); when it comes from TR, the frequency drifting rate should be relatively very high in the moderate frequency range (1.4 - 140 GHz); and when it comes from the corona, the frequency drifting rate should become relatively slow in the frequency range (100 MHz - 5.6 GHz). Different emission mechanism related to the heating processes will also produce different spectral features of the radio bursts. Such spectral features include frequency bandwidth, lifetime, frequency drifting rate, and polarization degree, etc. It is possible to identify and verify the heating mechanisms at different layers of the solar atmosphere through observations of radio ultra-wide dynamic spectrometer during quiet Sun.

3. The non-thermal radio emissions related to the solar atmospheric heating mechanisms span an ultra broadband frequency range from meter waves (down to 100 MHz) to millimeter waves (up to more than 100 GHz), which require the observations should have high sensitivity, high temporal resolution, high frequency resolution, and also a combination of spectral and imaging observations simultaneously during the period of quiet Sun, without the effects of solar powerful eruptions.

\begin{acknowledgements}
This work is supported by the Strategic Priority Research Program of the Chinese Academy of Sciences XDB0560000, the National Key R\&D Program of China 2021YFA1600503, 2022YFF0503001, 2022YFF0503800, the National Natural Science Foundation of China (NSFC) No.12173050, and the International Partnership Program of the Chinese Academy of Sciences 183311KYSB20200003. We also acknowledge the use of data from the Chinese Meridian Project.
\end{acknowledgements}

\label{lastpage}


\begin{thebibliography}{99}

\bibitem[Alfven(1947)]{Alfven 1947} Alfven, H. 1947, MNRAS, 107, 211

\bibitem[Alissandrakis(2019)]{Alissandrakis 2019} Alissandrakis, C. E., Bouratzis, C., Hillaris, A. 2019, A$\&$A, 627, A133

\bibitem[Aschwanden(1997)]{Aschwanden 1997} Aschwanden, M. J., Benz, A. O. 1997, ApJ, 480, 825

\bibitem[Aschwanden(2012)]{Aschwanden 2012} Aschwanden, M. J., Freeland, S. L. 2012, ApJ, 754, 112

\bibitem[Asgari-Targhi(2013)]{Asgari-Targhi 2013} Asgari-Targhi, M., Van Ballegooijen, A. A., Cranmer, S.R., et al. 2013, ApJ, 773, 111

\bibitem[Bastian(1998)]{Bastian 1998} Bastian, T. S., Benz, A. O., Gary, D. E. 1998, ARA$\&$A, 36, 131

\bibitem[Berghmans(2021)]{Berghmans 2021} Berghmans, D., Auchere, F., Long, D. M., et al. 2021, A$\&$, 656, L4

\bibitem[Bouratzis(2019)]{Bouratzis 2019} Bouratzis, C., Hillaris, A., Alissandrakis, C. E., et al. 2019, A$\&$A, 625, A58

\bibitem[Bradshaw(2015)]{Bradshaw 2015} Bradshaw, S. J., Klimchuk, J. A. 2015, ApJ, 811, 129

\bibitem[Chen(2015)]{Chen 2015} Chen, B., Bastian, T. S., Shen, C. C., et al. 2015, Sci., 350, 1238

\bibitem[Chen(2021)]{Chen 2021} Chen, Y., J., Przybylski, D., Peter, H., et al. 2021, A$\&$A, 656, L7

\bibitem[Chen(2022)]{Chen 2022} Chen, J., Erdelyi R, Liu J.J., et al. 2022, Front. Astron. Space Sci, 8, 786856

\bibitem[Chitta(2023)]{Chitta 2023} Chitta, L., P., Zhukov, A. N., Berghmans, D., et al. 2023, Sci., 361, 867

\bibitem[Choudhuri(1993)]{Choudhuri 1993} Choudhuri, A. R., Dikpati, M., Banerjce, D. 1993, ApJ, 413, 811

\bibitem[Crosby(1993)]{Crosby 1993} Crosby, N. B., Aschwanden, M. J., Dennis, B. R. 1993, SoPh, 143,
275

\bibitem[Cranmer(2007)]{Cranmer 2007} Cranmer, S. R., Van Ballegooijen, A. A., Edgar, R., J. 2007, ApJS, 171, 520

\bibitem[Cranmer(2015)]{Cranmer 2015} Cranmer, S. R., Woolsey, L., N. 2015, ApJ, 812, 71

\bibitem[Davila(1987)]{Davila 1987} Davila, J. 1987, ApJ, 317, 514

\bibitem[Dulk(1985)]{Dulk 1985} Dulk, G.A. 1985, ARA$\&$A, 23, 169

\bibitem[De Pontieu(2004)]{De Pontieu 2007} De Pontieu, B., Erdelyi, R., James, S. P. 2004, Nature, 430, 29

\bibitem[De Pontieu(2007)]{De Pontieu 2007} De Pontieu, B., McIntosh, S. W., Carlsson, M., et al. 2007, Sci., 318, 574

\bibitem[De Pontieu(2011)]{De Pontieu 2011} De Pontieu, B., McIntosh, S. W., Carlsson, M., et al. 2011, Sci., 331, 7

\bibitem[Gary(2023)]{Gary 2023} Gary, D.E. 2023, ARA$\&$A, 61, 427

\bibitem[Hashim(2021)]{Hashim 2021} Hashim, P., Hong, Z. X., Ji, H. S., et al. 2021, RAA, 21, 105

\bibitem[Heyvaerts(1983)]{Heyvaerts 1983} Heyvaerts, J., Priest, E. R. 1983, A$\&$A, 117, 220

\bibitem[Hou(2023)]{Hou 2023} Hou, Z. Y., Tian, H., Su, W., et al. 2023, ApJ, 953, 171

\bibitem[Hudson(1991)]{Hudson 1991} Hudson, H. S. 1991, SoPh, 133, 357

\bibitem[Jess(2009)]{Jess 2009} Jess, D. B., Mathioudakis, M., Erdelyi, R., et al. 2009, Sci, 323, 1582

\bibitem[Ji(2012)]{Ji 2012} Ji, H. S., Cao, W. D., Goode, P. R. 2012, ApJL, 750, L25

\bibitem[Ji(2021)]{Ji 2021} Ji, H.S., Hashim, P., Hong, Z. X., et al. 2021, RAA, 21, 179

\bibitem[Jiang(2015)]{Jiang 2015} Jiang, F. Y., Zhang, J., Yang, S. H. 2015, PASJ, 67, 40

\bibitem[Jiang(2024)]{Jiang 2024} Jiang, P., Chen, R.R., Gan, H. Q., et al. 2024, Astron. Tech. $\&$ Inst., 1, 84

\bibitem[Jin(2021)]{Jin 2021} Jin C.L., Zhou G.P., Wang J. X. 2021, ApJL, 914, L35

\bibitem[Kerr(2012)]{Kerr 2012} Kerr, R.A. 2012, Sci., 336, 1099

\bibitem[Klimchuk(2006)]{Klimchuk 2006} Klimchuk, J. 2006, SoPh, 234, 41

\bibitem[Klimchuk(2015)]{Klimchuk 2015} Klimchuk, J. 2015, Phil. Trans. R. Soc. A, 373, 20140256

\bibitem[Lee(2000)]{Lee 2000} Lee, L.C., Wu, B.H. 2000, ApJ, 535, 1014

\bibitem[Leonardo(2023)]{Leonardo 2023} Leonardo, D.J.S., Fidel, C. 2023, Phys. Today, 76, 34

\bibitem[Lu(2024)]{Lu 2024} Lu, Z.K., Chen, F., Ding, M.D., et al. 2024, Nature Astron., 8, 706

\bibitem[Moore(1991)]{Moore 1991} Moore, R. L., Musielak, Z. E., Suess, S. T., An, C. H. 1991, ApJ, 378, 349

\bibitem[Mason(2023)]{Mason 2023} Mason, J. P., Werth, A., West, C. G., et al. 2023, ApJ, 948, 71

\bibitem[Narain(1996)]{Narain 1996} Narain, U., Ulmschneider, P. 1996, SSRv, 75, 563

\bibitem[Parker(1988)]{Parker1988} Parker, E., N. 1988, ApJ, 330, 474.

\bibitem[Parnell(2000)]{Parnell2000} Parnell, C., E., Jupp, P.E. 2000, ApJ, 529, 554.

\bibitem[Rappazzo(2007)]{Rappazzo2007} Rappazzo, A. F., Velli, M., Einaudi, G., et al. 2007, ApJL, 657, L47.

\bibitem[Red(2014)]{Red2014} Red, H. A. S., Ratcliffe, H. 2014, RAA, 14, 773.

\bibitem[Samanta(2019)]{Samanta2019} Samanta, T., Tian, H., Yurchyshyn, V., et al. 2019, Sci, 366, 890.

\bibitem[Schwarzschild(1948)]{Schwarzschild1948} Schwarzschild, M., 1948, ApJ, 107, 1

\bibitem[Singh(2015)]{Singh2015} Singh, N., 2015, ApJL, 810, L1

\bibitem[Sturrock(1999)]{Sturrock1999} Sturrock, P.A., 1999, ApJ, 521, 451

\bibitem[Shimizu(1995)]{Shimizu1995} Shimizu, T., 1995, PASJ, 47, 251

\bibitem[Su(2015)]{Su2015} Su, W., Cheng, X., Ding, M.D., et al. 2015, ApJ, 804, 88

\bibitem[Tan(2013)]{Tan2013} Tan, B.L., 2013, ApJ, 773, 165.

\bibitem[Tan(2014)]{Tan2014} Tan, B.L., 2014, ApJ, 795, 140.

\bibitem[Tan(2016)]{Tan2016} Tan, B. L, Meszarosova H, Karlicky M, et al. 2016, ApJ, 819, 42.

\bibitem[Tan(2019)]{Tan2019} Tan, B. L, Chen, N.H., Yang, Y. H., et al. 2019, ApJ, 885, 90.

\bibitem[Tan(2020)]{Tan2020} Tan, B.L., Yan, Y., Li, T., et al. 2020, RAA, 20, 90.

\bibitem[Tan(2024)]{Tan2024} Tan, B.L., Huang, J., Zhang Y., et al. 2024, Universe, 10, 82.

\bibitem[Testa(2014)]{Testa2014} Testa, P., De Pontieu, B., Allred, J., et al. 2014, Sci, 346, 1255724.

\bibitem[Tian(2014)]{Tian2014} Tian, H., DeLuca, E. E., Cranmer, S. R., et al. 2014, Sci, 346, 1255711.

\bibitem[Van Doorsselaere(2020)]{Van Doorsselaere2020} Van Doorsselaere, T., Srivastava, A. K., Antolin, P., et al. 2020, SSRv, 216, 140

\bibitem[Van Ballegooijen(2017)]{Van Ballegooijen2017} Van Ballegooijen, A., A., Asgari-Targhi, M., Voss, A. 2017, ApJ, 849, 46

\bibitem[Vernazza(1981)]{Vernazza1981} Vernazza, J. E., Avrett, E. H., Loeser, R., et al. 1981, ApJS, 45, 635

\bibitem[Wager(1983)]{Wager1983} Wager, W. J., MacQueen, R. M.  1983, A$\&$A, 120, 136

\bibitem[Walsh(2003)]{Walsh2003} Walsh, R.W., Ireland, J. 2003, A$\&$ARv, 12, 1

\bibitem[Wan(2021)]{Wan2021} Wan, J. L., Tang, J. F., Tan, B. L., et al.  2021, A$\&$A, 653, A38

\bibitem[Withbroe(1977)]{Withbroe1977} Withbroe, G.L., Noyes, R.W. 1977, ARA$\&$A, 15, 363

\bibitem[Yan(2021)]{Yan2021} Yan, Y.H., Chen, Z. J., Wang, W., et al. 2021, Frontiers Astron. Space Sci., 8, 20

\bibitem[Yu(2020)]{Yu2020} Yu, S.J., Chen, B., Reeves, K.K., et al. 2020, ApJ, 900, 17.

\bibitem[Yuan(2023)]{Yuan2023} Yuan, D., Fu, L. B., Cao, W.D., et al. 2023, Nature Astron., 8, 856.

\bibitem[Zhang(2011)]{Zhang2011} Zhang, J., Liu, Y. 2011, ApJL, 741, L7.

\end{thebibliography}
\end{document}